\documentclass[aps,prl,twocolumn,a4paper,10pt]{revtex4-1}

\usepackage[T1]{fontenc}		
\usepackage[english]{babel}		
\usepackage[latin1]{inputenc}	
\usepackage{times}

\usepackage{amsmath}			
\usepackage{amssymb}			
\usepackage{bbm}				
\usepackage{mathtools}
\usepackage{gensymb}
\usepackage{simpler-wick}
\usetikzlibrary{tikzmark}
\DeclarePairedDelimiterX\Dirbraket[3]{\langle}{\rangle}%
{#1\,\delimsize\vert\,\mathopen{}#2\,\delimsize\vert\,\mathopen{}#3}

\usepackage[colorlinks=true, a4paper=true, pdfstartview=FitV,linkcolor=blue, citecolor=blue, urlcolor=blue]{hyperref}

\usepackage{graphicx}			
\usepackage{graphics}			
\DeclareGraphicsExtensions{.pdf}
\usepackage{color}		

\newcommand{\bea}{\begin{eqnarray}}
\newcommand{\eea}{\end{eqnarray}}

\newcommand{\bk}{\mathbf{k}}

\begin{document}

\title{In situ controllable magnetic phases in doped twisted bilayer transition-metal dichalcogenides}

\author{Johan~Carlstr\"om }
\affiliation{Department of Physics, Stockholm University, 106 91 Stockholm, Sweden}
\date{\today}

\begin{abstract}
We study the electronic structure of hole-doped transition metal dichalcogenides for small twist-angels, where the onsite repulsion is extremely strong. Using unbiased diagrammatic Monte Carlo simulations, we find evidence for magnetic correlations which are driven by delocalization, and can be controlled in situ via the dielectric environment.
For weak spin-orbit coupling we find that the moderately doped system becomes anti-ferromagnetic, whilst the regime of strong spin-orbit coupling features ferromagnetic correlations. 
We show that this behavior is accurately predicted by an analytical theory based on moment expansion of the Hamiltonian, and analysis of corresponding particle trajectories. 
\end{abstract}
\maketitle


Moiré materials--resulting from the incommensurate stacking of single atomic layers--have emerged as an important platform for exploring strongly correlated physics. This field was initiated by the discovery of superconductivity \cite{Cao2018} and strongly correlated phases \cite{Cao2018SC} in twisted bilayer graphene (TBG) at "magic angles" where virtually flat bands occur in the spectrum \cite{Bistritzer12233}. While TBG remains a very active research field, it presents certain problems due to its theoretical complexity and the fragile nature of the electronic structure. 
TBG defies the construction of symmetric Wannier functions, meaning that it cannot be accurately described by a conventional lattice model \cite{PhysRevX.8.031089}, and this complicates theoretical analysis. Experiments are faced with the problem that the interesting correlated phases of TBG only exist for certain twist-angles, thus requiring extensive fine tuning. 

The twisted bilayer transition metal dichalcogenides (tTMD) have been identified as an attractive alternative to TBG \cite{Zhang2020} due being robust, highly tunable via the twist-angle and the dielectric environment, and relatively simple to model theoretically \cite{PhysRevLett.121.026402}.
To the first order approximation they can be described by a spin-orbit coupled Hubbard model on the triangular lattice. The bandwidth of the tTMD's can be tuned over several orders of magnitude by the twist angle, while spin-orbit coupling and doping can be controlled in situ \cite{PhysRevResearch.2.033087}. The electronic properties also evolve continuously over an extensive range of twist-angels, allowing for greater experimental control.

Experiments on the tTMD's have revealed a rich phenomenology, and the combination of a high degree of control and a simple theoretical description suggests that a more comprehensive understanding of strongly correlated physics may now be possible. In particular, several enigmatic phenomena observed in the cuprate superconductors have been reproduced recently.
At a $4.2\degree$ twist-angle, a strange metal regime has been identified at small doping, giving way to a fermi liquid at higher carrier concentrations \cite{2021QC}. 
At a $5.1\degree$ twist, superconductivity has been observed with a critical temperature of $T_c\approx 3K$ \cite{Wang2020} . 
These twist angles correspond to a parameter regime where the onsite repulsion is comparable to the bandwidth \cite{PhysRevResearch.2.033087,zang2021dynamical}, implying strong magnetic correlations.  
For physically realizable model parameters, super-exchange processes are predicted to favor anti-ferromagnetism \cite{PhysRevResearch.2.033087,PhysRevB.104.075150}, and so this scenario bears a striking similarity to the cuprate superconductors \cite{RevModPhys.78.17}.
Recently, strange metal behavior was also predicted theoretically using dynamic mean-field theory \cite{zang2021dynamical}. 

Thus far, both theory and experiments have focused on the scenario of a large twist angle where the bandwidth is comparable to the onsite repulsion. At half-filling or small doping, this results in super-exchange processes with an energy scale that is comparable to the hopping integral. At the smallest twist-angles, this situation changes dramatically as the onsite repulsion becomes much larger than the bandwidth, effectively suppressing super-exchange. 

In this paper we focus on magnetism in the regime of small twist angles. Using diagrammatic Monte Carlo \cite{Van_Houcke_2010}, we establish that delocalization of the charge carriers can induce either ferromagnetism or anti-ferromagnetism, depending on the spin-orbit coupling, which is in turn controlled by the dielectric environment. 

{\it Model---}The low-energy physics of the tTMD's can approximately be described by a generalized "Moiré Hubbard model" on the triangular lattice with spin-orbit coupling \cite{PhysRevResearch.2.033087,PhysRevLett.121.026402}:
\bea
H\!=\!-\!\sum_{\sigma,\langle ij\rangle} t_{ij\sigma} c_{\sigma,i}^\dagger c_{\sigma,j} \!+\!\sum_i U_0 \hat{n}_{\downarrow,i} \hat{n}_{\uparrow,i},\;
t_{ij\sigma}\!=\!|t| e^{i\sigma\phi_{ij}}.\;\;\;\;\label{Hubbard}
\eea
Here, $\sigma$ refers to both spin and valley, as these are locked in transition metal dichalcogenides \cite{PhysRevLett.108.196802,Bawden2016}. The phase $\phi$ is determined by the potential difference between the two layers, and is thus controllable in situ. DFT calculations indicate that realistically achievable parameters correspond to approximately $0\le \phi\le \pi/3$ \cite{Wang2020}. 
In this model, interchanging the spin corresponds to taking $\phi\to -\phi$, while particle-hole transformation is equivalent to taking $\phi\to\phi+\pi$.

The hopping integral $|t|$ decays exponentially with the moiré period, which is in turn determined by the twist angle $\theta$. As a result, the bandwidth can be tuned over several orders of magnitude in the interval $1\degree\le\theta\le 5\degree$ \cite{PhysRevResearch.2.033087}. By contrast, the onsite repulsion changes by less that a factor $2$ in the same interval, thus permitting the relative interaction $U/t$ to change dramatically.  
At $\theta=4-5\degree$, the hopping integral has been estimated to $|t|\sim 100$K while $U/t\sim 8$ \cite{zang2021dynamical}, and so the regime $\theta\approx 1\degree$ corresponds to extremely large onsite interactions with a corresponding suppression of super-exchange processes. 
Higher order corrections to the model (\ref{Hubbard}) consist of longer range hopping and also non-local repulsive interactions. 
 
In this work, we focus on the physics corresponding to small twist-angles, where $U$ is much greater than the bandwidth. For our calculations, we will specifically consider the strong coupling limit $U\to\infty$, though our results remain valid for a range of finite couplings as well. 
In this regime, magnetic correlations emerge as the effective density of states of a charge carrier is renormalized by the mean-free path \cite{brink}.
 On bipartite lattices, the lowest energy for a single charge carrier is obtained on a polarized background, leading to the well known Nagaoka theorem \cite{PhysRev.147.392}, with generalizations to finite doping as well \cite{PhysRevLett.108.126406}.

For the Moiré Hubbard model, no analytical results exist in Nagaoka's scenario, though certain qualitative assessments can be made based on the moments of the kinetic energy. First, we note that the moments of the dispersion $\epsilon_\bk$ can be connected to an expansion of $H$ for a state consisting of a single localized hole on a polarized background (denoted by $\psi$),
\bea
\int d\bk \epsilon_\bk^n\sim
 \langle \psi |H^n |\psi\rangle.\label{moments}
\eea
Here, the expansion in $H$ can be regarded as a quantum walk, with non-vanishing contributions resulting when the system returns to the initial state. For a polarized background, the mean-free path is infinite, and the dopant is correspondingly an ideal fermion. By contrast, other types of backgrounds lead to a reduced mean-free path that renormalize the effective density of states seen by the carrier \cite{brink}. 

\begin{figure}[!htb]
\includegraphics[width=\linewidth]{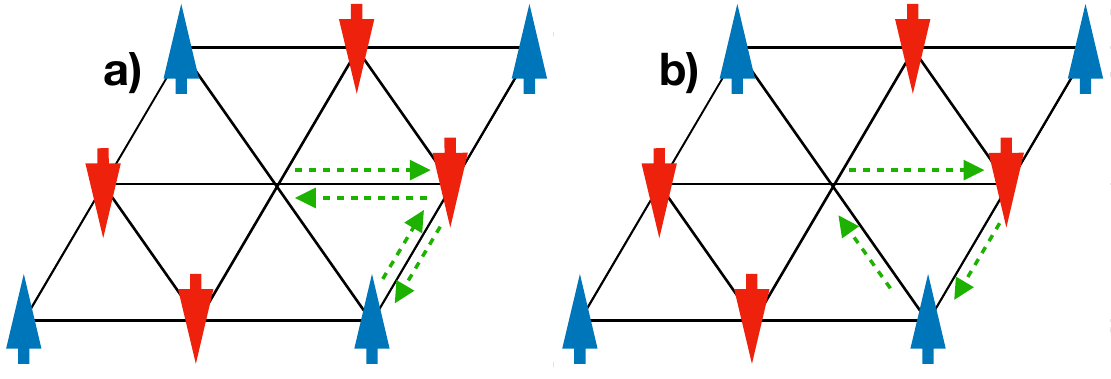}
\caption{
{\bf Two classes of trajectories.} The self-retracing path (a) is a zero-area loop that inherently preserves the spin-background. In the non-retracing path (b), the carrier exchanges elements of the background, meaning that the interference of the initial and final state becomes dependent on the spin-background. 
}
\label{traj}
\end{figure}

There are two principal types of trajectories in the quantum walks (\ref{moments}): Self-retracing paths that inherently preserve the spin-background, and non-retracing paths that exchange elements of the background, see also Fig. \ref{traj}. It is only the latter of the two that depends on the spin-background. Since all odd terms in the expansion (\ref{moments}) must enclose a non-vanishing area, it follows that they are non-retracing. Hence, all odd terms in the expansion (\ref{moments}) are significantly reduced if the mean-free path is finite, as occurs in the absence of ferromagnetic correlations. Furthermore, most of the effective bandwidth of a carrier on a Mott insulating background is contained in the self-retracing paths \cite{brink,blomquist2019ab}. 

From the preceding considerations, we arrive at the following conjecture for the interplay of delocalization and magnetic correlations: When the system is weakly to moderately hole-doped, the charge carriers will occupy the top of the effective band and provide a kinetic energy that is $dE/N \approx -\epsilon_{max}$, where $N$ is the number of carriers and $\epsilon_{max}$ is the effective band top. If the odd terms in (\ref{moments}) increase the band top $\epsilon_{max}$, then delocalization will drive drive the system towards ferromagnetism, whilst if they decrease the band top, then anti-ferromagnetic correlations will be preferred, as this minimizes the mean-free path.

\begin{figure}[!htb]
\includegraphics[width=\linewidth]{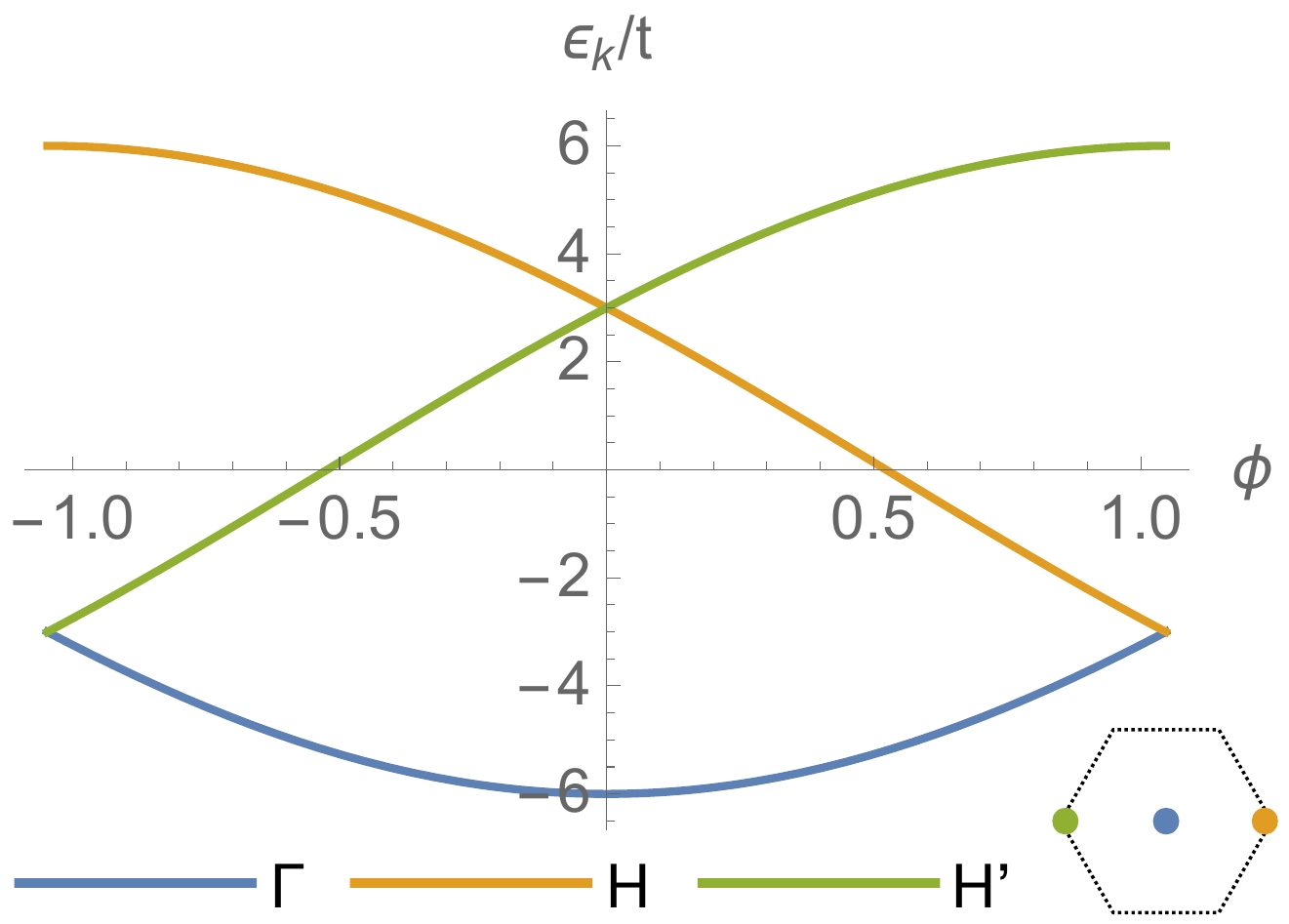}
\caption{
{\bf Kinetic energy} at $\bk=(0,0)$ and $\bk=\pm (4\pi/3,0)$ for $|\phi|\le \pi
/3 $ and $\sigma=1$. At the edge of the parameter region ($\phi=\pm \pi/3$), the band top is situated at $\epsilon_k=6t$ while the bottom corresponds to $-3t$. This asymmetry results from odd terms in (\ref{moments}), and allows a hole to delocalize with an energy of $-6t$ on a ferromagnetic background, providing a strong case for ferromagnetism. At $\phi=0$, the situation is the precise opposite.  
}
\label{band}
\end{figure}

Fig. \ref{band} shows how the band top and bottom depend on the spin orbit coupling. At the edge of the parameter space ($\phi=\pm \pi/3$) the band top is situated at $6t$, while the bottom corresponds to $-3t$, implying that the odd terms in (\ref{moments}) increase $\epsilon_{max}$. At $\phi=0$, this situation is reversed. Based on the preceding conjecture, this implies that the system becomes ferromagnetic at the edges of the parameter space, and antiferromagnetic at the origin, with a crossover occurring at $\phi_c\approx \pm \pi/6$.

Because of spin-orbit coupling, SU(2) symmetry is explicitly broken down to Z$_2$--corresponding to time-reversal--when $\phi\not=n\pi$. This means that the Mermin-Wagner theorem \cite{PhysRevLett.17.1133} does not apply, and that a long-range ferromagnetic order may persist to finite temperatures, despite that the system is 2D.

{\it Numerical treatment---} To the test the preceding conjecture on how delocalization interacts with magnetism, we employ diagrammatic Monte Carlo, which is a numerical protocol based on stochastic sampling of Feynman type graphs \cite{Van_Houcke_2010}. This method has the advantage that it can be employed directly in the macroscopic limit, and is  asymptotically exact, as the only systematic source of error is truncation of the series. 
To be able to address systems with strong interactions we use a particular formulation known as strong-coupling diagrammatic Monte Carlo (SCDMC) \cite{PhysRevB.103.195147,0953-8984-29-38-385602,PhysRevB.97.075119,carlstrom2021spectral}, where the diagrammatic elements are connected vertices of propagating electrons that are non-perturbative in $U$. The computational protocol employed here is outlined in detail in \cite{PhysRevB.103.195147}.

The expansion parameter is the hopping integral $t$. The principal observable that we compute is the polarization operator of the hopping integral, which we denote $\Pi_t(\omega,\bk)$. The dressed hopping integral is then obtained via the Bethe-Salpiter equation according to
\bea
\tilde{t}(\omega,\bk)=\frac{1}{t^{-1}(\bk)-\Pi_t(\omega,\bk)}.
\eea 
By conducting a skeleton expansion in the dressed hopping integral $\tilde{t}$, and iterating until convergence, we obtain a self-consistent solution that implicitly takes into account certain classes of diagrams up to infinite order. The dressed Greens function can be derived from the polarization according to 
\bea
G(\omega,\bk)=\frac{1}{\Pi_t^{-1}(\omega,\bk)-t(\bk)}.
\eea

Because SU(2) symmetry is explicitly broken, we expect the onset of ferromagnetism to be accompanied by critical behavior, which implies a divergent susceptibility with respect to symmetry-breaking perturbations.  
Furthermore, even in the proximity of a phase transition, the correction from diagrams at higher order tend to be large. This presents a challenge for diagrammatic methods, which are based on the expansion of an analytical function. 
To overcome this difficulty, we impose a strong symmetry-breaking perturbation on the system, in the form of an external field 
\bea
\delta H= B \frac{n_\uparrow-n_\downarrow}{2},\;B=B_0 t/\beta, \label{extfield}
\eea
where $t$ is the bare hopping integral, and $\beta$ is the inverse temperature. At half-filling or in the atomic limit, the response to the perturbation (\ref{extfield}) is that of a paramagnet, so that
\bea
\rho_{min}=\frac{n_\uparrow}{n_\uparrow+n_\downarrow}=\frac{e^{-B_0/2}}{e^{-B_0/2}+e^{B_0/2}},\label{paramagn}
\eea
which defines the relative density of the minority component, $\rho_{min}$. We choose $B_0\approx2.946$, so that $\rho_{min}=0.05$ for a paramagnet. 
Next, we consider hole-doping the system. We choose our chemical potential such that the carrier density is $20\%$ when the system is fully polarized, though the resulting equation of state will depend on renormalization of the kinetic energy due to a finite mean-free path in the system. Now, we conduct a self-consistent expansion in $t$ to obtain the Greens function.

\begin{figure*}[!htb]
 \hbox to \linewidth{ \hss
\includegraphics[width=\linewidth]{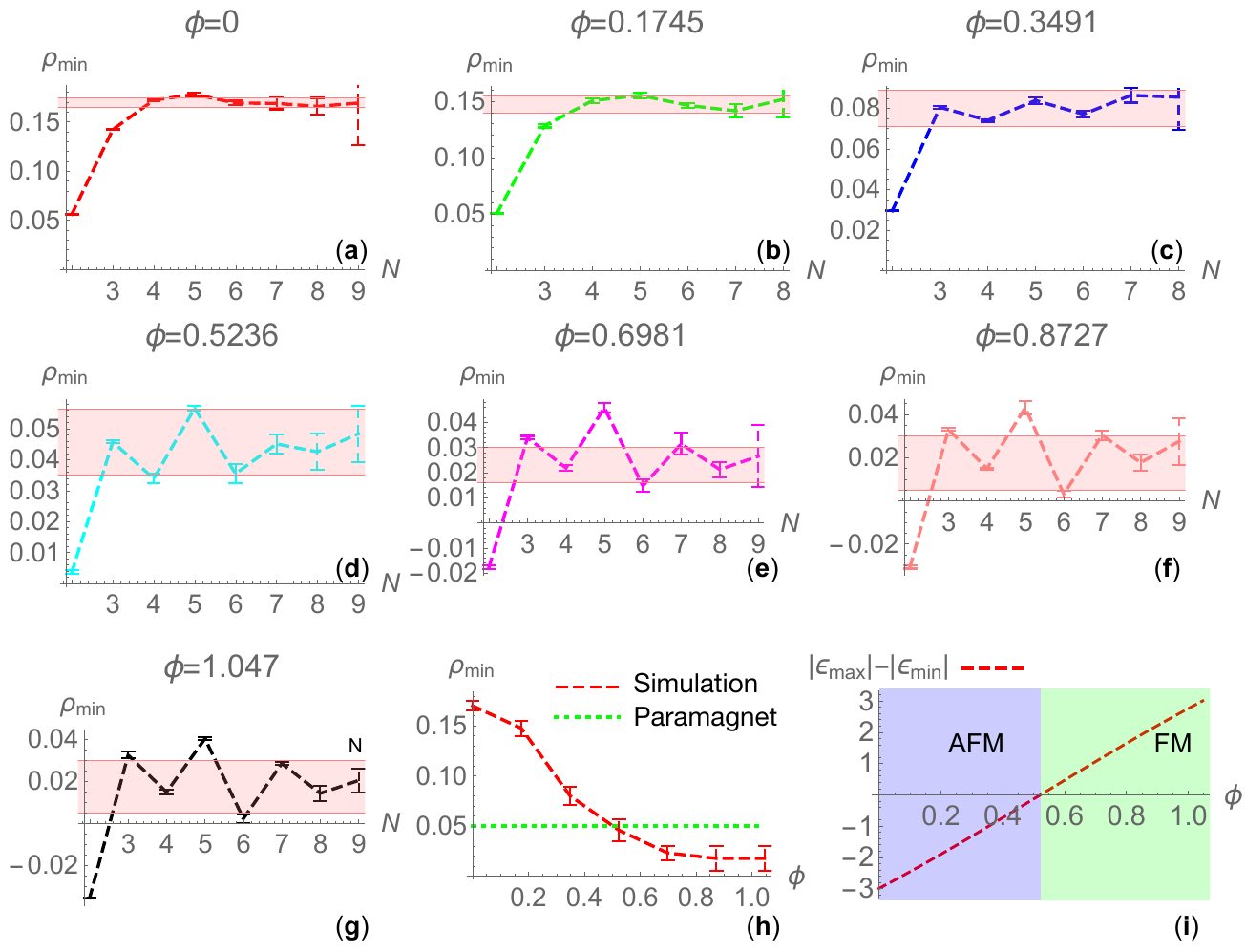}
 \hss}
\caption{
{\bf Relative density of the minority component} ($\rho_{min}=n_\uparrow/(n_\uparrow+n_\downarrow)$) obtained from diagrammatic Monte Carlo expansion up to an order $N=8$ or $N=9$ (a-g). The expansion is conducted for spin-orbit couplings in the interval $0\le \phi\le \pi/3$ and in a strong external field according to (Eq. \ref{extfield}).
The external field is chosen such that $\rho_{min}=0.05$ in the atomic limit (see Eq. \ref{paramagn} ), and the deviation from this figure is thus a result of delocalization of charge carriers. 
The system is hole-doped, and the chemical potential in (a-g) is chosen such that the carrier density is $20\%$ if the system is fully polarized. 
The shaded region is an estimate for $\rho_{min}$ as $N\to\infty$. The results in (a-g) are summarized and compared to the paramagnetic response in (h). At small spin-orbit coupling, we observe an increase in the minority component, implying that delocalization favors a short mean-free path, and thus anti-ferromagnetism. At large spin-orbit coupling, the preference is for ferromagnetism. The critical point separating these regions can be identified by the crossing of the Monte Carlo data (red dashed line) and the paramagnetic response (green dotted line), which occurs at $\phi_c\approx 0.5$. 
We may compare this result to the initial conjecture: We expect that the type of magnetism favored by delocalization is primarily determined by the magnitude difference between the band top ($\epsilon_{max}$) and bottom ($\epsilon_{min}$). Correspondingly, we expect the critical point to occur at $\phi_c=\pi/6\approx 0.52$. The predicted phase diagram shown in (i) is thus in excellent agreement with our findings. 
}
\label{CMSIpanel}
\end{figure*}

In Fig. \ref{CMSIpanel} we display the relative density of the minority component for values of the spin-orbit coupling in the interval $0\le\phi\le \pi/3$ and an inverse temperature $\beta t=3$. In the atomic limit we expect $\rho_{min}=0.05$, and so a deviation from this figure is a result of delocalization. For small values of $\phi$, we find that the relative density of the minority component increases, implying that a short mean-free path decreases the kinetic energy of charge carriers, thus driving the system to form anti-ferromagnetic correlations. 
At strong spin-orbit coupling, we instead see a reduction of the minority component density, implying a preference for a long mean-free path and thus ferromagnetism. The crossover occurs at $\phi_c\approx 0.5$ which is very close to our prediction that $\phi_c=\pi/6\approx0.52$.

\begin{figure}[!htb]
\includegraphics[width=\linewidth]{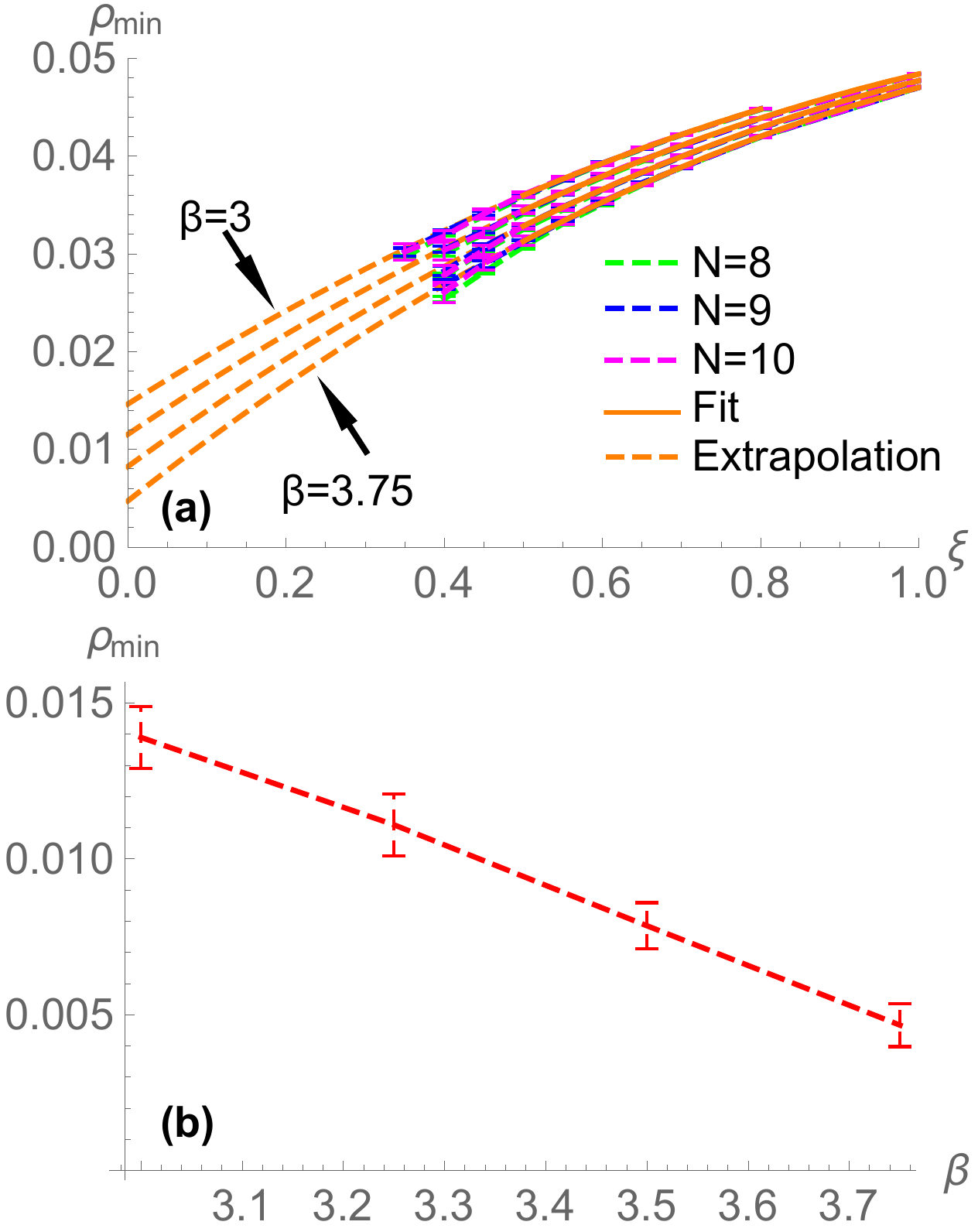}
\caption{
{\bf Temperature dependence} of the relative density of the minority component ($\rho_{min}=n_\uparrow/(n_\uparrow+n_\downarrow)$), obtained by resummation of the series. By reweighting diagram topologies according to Eq. (\ref{resummation}), we obtain a series for $\rho_{min}(N,\xi)$. We computed  $\rho_{min}(N,\xi)$ for $N=8,9,10$ and a range of resummation parameters $\xi$ (a). For $\xi\ge \xi_{min}\sim 0.45\;-\;0.5$, the solutions for different expansion orders $N$ coincide, indicating that the series for $\rho_{min}(N,\xi)$ has converged to $\rho_{min}(N\to\infty,\xi)$. Thus, we fit a second order polynomial to the simulation data in the region $\xi\ge \xi_{min}$ where $\rho$ has converged with respect to $N$, indicated by the solid orange line. The polynomial provides an extrapolation to $\xi=0$, indicated by the dashed orange line. We extracted these results for inverse temperatures $3\le\beta t\le 3.75$. 
The estimate for $\rho(N\to\infty,\xi=0)$ is summarized in (b). As the temperature decreases, the relative density of the minority component drops. It should be noted that the applied field is taken to be $B=B_0 t/\beta$, and so the decrease of the minority component density is only driven by the increasing energy scale of delocalization as compared to temperature. The system considered is hole-doped at $\sim 20\%$. 
}
\label{panel2}
\end{figure}

 Next, we focus on the point $\phi=\pi/3$, where we expect, and also find, the largest reduction of the minority component density. In Fig. \ref{CMSIpanel} (g) we observe that $\rho_{min}$ is oscillating with an amplitude that falls off rather slowly, suggesting that we are close to the convergence radius of the expansion. To reach lower temperatures we therefore employ a resummation scheme of the form 
\bea
W_i(\{x\})\to W_i(\{x\})e^{-\xi N_i}.\label{resummation}
\eea
Here, $W_i(\{x\})$ denotes the weight of the topology $i$ with internal variables $\{x\}$, while $N_i$ is the expansion order of this topology, and $\xi$ is the resummation parameter. From the reweighted series we obtain observables which formally depend on the expansion order and the resummation parameter $\xi$, so that
\bea
\rho_{min}=\rho_{min}(N,\xi).\label{reO}
\eea
For a series with a finite convergence radius we expect that $\rho_{min}(N,\xi)$ converges as $N\to\infty$ for a range $\xi\ge  \xi_{min}$. Our aim is now to estimate $\rho_{min}(N\to\infty,\xi\ge \xi_{min})$ and extrapolate this result to $\xi=0$. 
Thus, we compute $\rho_{min}(N,\xi)$ for $N=8,9,10$ and a range of $\xi$. Fig. \ref{panel2} (a) shows $\rho_{min}$ as a function of the resummation parameter for inverse temperatures $3\le\beta\le 3.75$. For $\xi\ge \xi_{min}\sim 0.45\;-\;0.5$ the lines corresponding to different expansion orders $N$ coincide, indicating convergence. Thus, we fit a second order polynomial to the simulation data in the region $\xi\ge \xi_{min}$. Extrapolating this function to $\xi=0$ gives an estimate for $\rho_{min}(N\to\infty,\xi=0)$. A summary of the estimate is given in Fig. \ref{panel2} (b): As the temperature is reduced, $\rho_{min}$ decreases monotonically. It should be stressed, that this calculation is conducted for a magnetic field that decreases with temperature according to Eq. (\ref{extfield}), so that the reduction of the minority density is entirely the result of delocalization. 

The error bars in Fig. \ref{panel2} (b) were obtained by varying the cutoff $\xi_{min}$ and observing how the result changes. However, it should be stressed that there is also an implicit error associated with the choice of fitting function. We also tried other fitting functions, including a third order polynomial, though this resulted in greater sensitivity to noise in the simulation data. Nonetheless, the key qualitative feature of the series--that $\rho_{min}$ decreases as the temperature is reduced--is reproduced for all fitting functions we tried.



{\it Discussion---}
We have identified a mechanism by which delocalization of charge carriers gives rise to magnetic correlations in the moiré Hubbard model. Based on analysis of particle paths, we conjecture a cross over from anti-ferromagnetic correlations at weak spin-orbit coupling to a ferromagnetic regime at strong coupling. Comparison of this theory to diagrammatic Monte Carlo simulations show excellent agreement. As the temperature is reduced, we confirm a rapid reduction of the minority component in the strong-coupling regime, indicating that the system eventually becomes fully polarized. 

While the results reported here were obtained in the strong-coupling limit, they remain relevant for physical parameter regimes of the TTMD's. For example, at twist-angles of $\theta =4-5 \degree$, estimates of effective model parameters yield $|t|\sim 100$K and $U/t\sim 8$ \cite{zang2021dynamical}. If the hopping integral is shrunk by two orders of magnitude via the twist-angle, and the onsite repulsion changes by less than a factor 2 \cite{PhysRevResearch.2.033087}, then this gives a super-exchange $J/t\sim 4t/U\sim 100$ and $t\sim 1$K. The energy scale of a magnetic bond is thus $\epsilon_{M}\sim t/400$, implying that $\epsilon_{M}\beta\sim 10^{-2}$ in our simulations. Thus, a conservative estimate in this scenario is that at the energy scale of magnetic bonds would be at least two orders of magnitude smaller than the simulated temperature, which corresponds to $0.25-0.33$K. This clearly demonstrates the existence of parameter regimes where super-exchange is irrelevant, and the strong coupling limit is an accurate description of the system. 

In the polarized regime, the moiré Hubbard model (\ref{Hubbard}) reduces to a system of ideal single-component fermions. However, if we also include non-contact interactions, then we obtain an effective "polarized moiré model" of the form
\bea
H=-\!\sum_{\langle ij\rangle} t_{ij} c_{i}^\dagger c_{j} \!+\!\sum_{ij} V_{ij} \hat{n}_{i} \hat{n}_{j},\;
t_{ij}\!=\!|t| e^{i\phi_{ij}}.\label{polarized}
\eea
Here, we have not included beyond nearest-neighbour hopping as these terms are extremely small, and most likely not important. The nearest neighbour interactions can be quite large however: An approximative assessment suggests that $U/V_{NN}\sim 15$ at a twist angle of $1\degree$  and $U/V_{NN}\sim 3$ at $5\degree$ \cite{PhysRevResearch.2.033087}. In our example, where $U/t\sim 400$, this would imply that $V_{NN}>25t$, so that the interaction is much larger than the bandwidth. Even beyond nearest neighbour interaction, the energy scale can be comparable to or larger than the bandwidth.

We expect nonlocal interactions to have limited impact on kinetically driven magnetism, since these terms do not affect the interaction between charge carriers and the spin-background. However, at much lower temperatures they may drive instabilities in the system. 

Finally, we note that the magnetic phases that appear for small twist-angles can be controlled via the dielectric environment. In the ferromagnetic regime, we expect that the minority component becomes gapped at sufficiently low temperature, implying that the system only transmits currents of the majority component. This would allow TTMD's to be used to construct "spin-filters" that are controllable in situ.


This work was supported by the Swedish Research Council (VR) through grant 2018-03882. Computations were performed on resources provided by the Swedish National Infrastructure for Computing (SNIC) at the National Supercomputer Centre in Linköping, Sweden.  The author would like to thank Ahmed Abouelkomsan and Emil Blomquist for important input and discussions.  

\bibliography{biblio.bib}

\end{document}